\documentstyle[]{l-aa}

\begin{document}
\input{psfig}

\thesaurus{10.07.2; 11.05.1; 11.09.1 NGC 5018}

\title{The globular cluster system of an ``old'' merger: NGC 5018
\thanks{Based on data collected at the European Southern Observatory,
	 La Silla, Chile}}

\author {Michael Hilker \inst{1} \& Markus Kissler--Patig \inst{2,1} 
} 

\offprints {M.~Hilker}

\institute{
Sternwarte der Universit\"at Bonn, Auf dem H\"ugel 71, 53121 Bonn, Germany
\and 
European Southern Observatory, Karl-Schwarzschild-Str.~2, 85748 Garching, Germany
}

\date {2.5.96}

\maketitle

\begin{abstract}
We study the properties of the globular cluster system of the ``old''
merger NGC 5018. We detect a relatively poor globular cluster system,
that can be divided into at least two populations based on colors:
a small population of
blue clusters, aged between several hundred Myr to 6 Gyr, the 
formation of which was probably induced by the last interaction in the galaxy
group; and one (possibly two) population(s) of 
old(er) clusters that can be associated with the original galaxy(ies).

This composite globular cluster system is as elliptical ($\epsilon=0.5\pm0.3$)
as the best oblate model for the luminous component (about 
E6), but as diffuse as the dark halo implied by the same model. In terms
of a power law slope, we find a fall-off of $-1.3\pm0.4$ for the globular
cluster density profile, while the intensity of the stellar light falls
off with $-2.2\pm0.2$.

We sample the globular cluster luminosity function down to $V=24.4$ mag, and 
do not reach the turnover. Fits to our data with Gaussians and $t_5$ functions
can only constrain the turn-over to $V_{\rm TO}=26.3\pm0.6$
mag. The total number of globular clusters derived from the luminosity
function and the surface density profile of the globular clusters  lies
then between $800\pm270$ and $1700\pm750$. The amount of young globular
clusters is estimated to be less than 10\% of the total, evidence against a
significant increase of globular clusters by the merger event in this
case.

The specific frequency, for which uncertainties in distance and total
number cancel, is $S=1.0\pm0.7$, lower than expected for a normal
elliptical galaxy. However, if the reported young luminous stellar
population would brighten NGC 5018 by one magnitude compared with normal
ellipticals, the corrected $S$ value would be about 3, comparable with values
found for ellipticals in small groups.

Finally, the study confirms that NGC 5018 contains a large amount of dust. 
The globular clusters might even reveal a dust concentration. 


\end{abstract}


\section{Introduction}

With the study of globular clusters in interacting and merging galaxies, 
globular cluster systems gain in importance for
the understanding of the evolutionary history of elliptical galaxies.
Properties like the luminosity function, specific frequency, morphology, and
color distribution of a globular cluster system contain information on 
the globular cluster formation history, as well as on the distance and history 
of the parent galaxy. Several hypotheses about the formation of globular
clusters and globular cluster systems are discussed in the literature 
(e.g.~Harris 1991, 1994 and Richtler 1995 for reviews, Ashman \& Zepf 1992 for
formation in mergers).  Besides primordial formation of globular clusters 
there are also indications that the
formation can be triggered by interactions between galaxies like merging
and/or accretion of dwarf galaxies. The study of ellipticals showing 
indications of merging events or interactions provides the possibility
to get a direct insight into the building up of globular cluster systems.

Examples of young globular clusters in recent merger galaxies
are becoming more numerous (e.g.~ NGC 3597: Lutz
1991; NGC 1275: Holtzman et al.~1992; NGC 7252: Whitmore et al.~1993, 
Schweizer \& Seitzer 1993; He2-10: Conti \& Vacca 1994; NGC 4038/4039,
Whitmore \& Schweizer 1995). 
NGC 5018, that underwent an interaction about 1 Gyr ago, might
be a similar system, probably even older than the previous examples. 
\begin{figure}
\psfig{figure=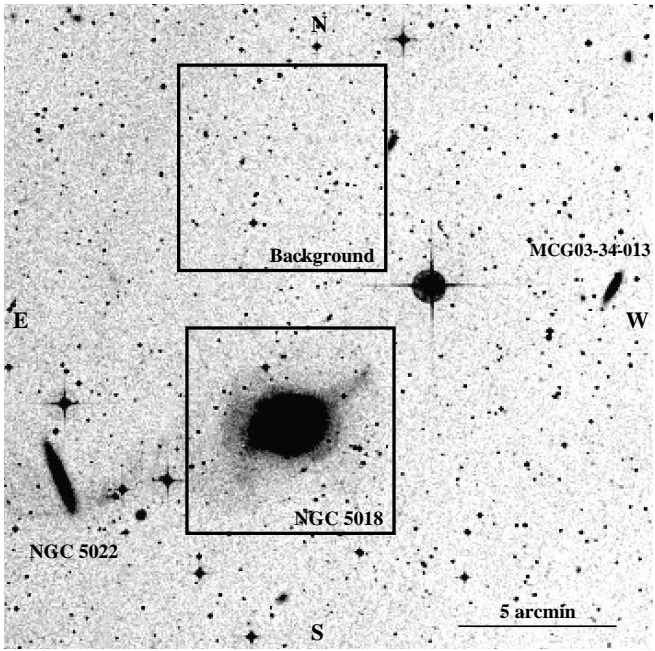,height=9cm,width=9cm
,bbllx=43mm,bblly=75mm,bburx=172mm,bbury=205mm}
\caption{Image taken from the Digital Sky Survey, showing the surroundings 
of NGC 5018}
\end{figure}

NGC 5018 is the brightest member of a small group of 5 galaxies (Gourgoulhon
et al.~1992). It is classified as a ``normal'' giant elliptical (E4 in the
RSA survey; E3 in the RC3 catalog) but shows several indications for 
a past merger event.
Some properties of NGC 5018 are summarized in Table 1.
\begin{table}
\caption{Properties of  NGC 5018, taken from the RC3 (de Vaucouleurs et
al.~1991), except for $(V-I)$ taken from Goudfrooij et al.~(1994b)}
\begin{center}
\begin{tabular}{l l}
\hline
Name & NGC 5018 \\
RA(2000) & 13 13 01 \\
DEC(2000) & $-$19 31 12\\
l, b & 309.9, 43.1 \\
type & E3\\
$V$ & $10.77\pm 0.13$ mag\\
$v_{\rm opt}$ & $2794 \pm 15~{\rm km~s}^{-1}$ \\
$(B-V)$ & 0.92 mag\\
$(V-I)_{\rm inner}$ & 1.05 mag\\
\hline
\end{tabular}
\end{center}
\end{table}

Both Carollo \& Danziger (1994) and Goudfrooij et al.~(1994b)
investigated in detail the photometric profile of NGC 5018 out to
$\simeq 100\arcsec$ and found
similar results. The color of the galaxy is very blue 
($(V-I)=1.05$, Goudfrooij et al.~1994b), and the color gradient is steeper
than normally observed in ellipticals. This led Carollo \& Danziger (1994)
to conclude that either the galaxy is very blue in the outer parts, or that
the whole galaxy is rather blue and obscured and reddened in the inner
part (where the boundary between inner and outer parts is at about
$15\arcsec$ radius). Our study excludes this inner part.
Both also agree that, despite a probably complicated dust
distribution in the inner region, the 
isophotes are disky at all
radii (note that this is not in contradiction with a merger event,
e.g.~Lima-Neto \& Combes 1995).

NGC 5018 is also peculiar in having one of the lowest nuclear magnesium
absorption-line strengths for its absolute magnitude (Schweizer et
al.~1990). However, if the Mg$_2$ index profile (Carollo \& Danziger 1994),
that shows a strong decrease down to solar values inside 7\arcsec{} from the 
center, is extrapolated to outer radii it reaches again values normal for the
luminosity of NGC 5018. This suggests the presence of a young
stellar population in the central region, diluting the Mg$_2$ index. In
that case NGC 5018 could be considered as a ``classical'' merger remnant.
This also is strengthend by the presence of prominent ripples and shells (e.g.
Schweizer 1987).
Schweizer \& Seitzer (1992) attribute to NGC 5018 one of the highest fine
structure parameters, making it one of the best merger candidates, and derive
a heuristic merger age from $UBV$ colors of $5\pm2$ Gyr.

Deep HI observations with the VLA by Guhathakurta et al.~(1989) show a large HI
bridge between NGC 5022 and MCG~03$-$34$-$013, the galaxies situated to the E
and NW of NGC 5018 (see Fig. 1).
The authors claim it to be the first direct
observational evidence for the formation of a shell system when an
elliptical galaxy merges with a cold disk system, in a (at least)
three-body encounter. They derive from dynamical arguments a relatively
recent last interaction, about 600 Myr ago, that produced the HI plume,
and an orbital time for the outermost shell of 300 Myr.

Dynamical investigations show that NGC 5018 has a relatively
high velocity
dispersion ($220~{\rm km~s}^{-1}$, Davies et al.~1987) but a flat velocity
dispersion gradient. To reproduce this flat gradient a massive, diffuse
dark halo is required which is half as flattened (about E3) as the best
fitting oblate dynamical model of the galaxy light (about E6, see Carollo et 
al.~1995, Carollo \&  Danziger 1994). Also the HI velocity data (Kim et
al.~1988) support the existence of an substantial amount of dark matter in the
galaxy group around NGC 5018.
%
%

\section{Observations and Reduction}

\subsection{The observations}

Deep Bessell $V$ (ESO \#~451) and Gunn $i$ (ESO \#~461) CCD images were 
taken in direct imaging mode 
with the Danish 1.54m telescope at the
European Southern Observatory, La Silla, Chile during the night of 23/24.4.1995
under photometric and good seeing conditions (between $0\farcs8$ and $1\farcs1$,
as measured from the FWHM of stellar images).
The CCD chip in use was a Tektronix chip (ESO \#28) $1024^2$ 
pixels; the scale is $0.38\arcsec$ per pixel, giving a field size of
$6.4\arcmin\times6.4\arcmin$.
We obtained respectively five 10 and five 12 min exposures in $V$ and 
$i$ centered on NGC 5018. Short exposures of 1 min were taken in both filters
to compare our photometry with the published aperture photometry of the
galaxy (Poulain \& Nieto 1994) and check the zero point of our calibration.
An additional background field, located $11\arcmin$ north of
NGC 5018 (see Fig. 1), was observed to get an estimation of the contribution
of background objects in the NGC 5018 field.
Bias- and sky-flatfield exposures were taken at the beginning
and end of the night. Landolt (1992) standard fields were observed throughout
the night for the photometric calibration.

\subsection{The reduction}

All the reductions were done in IRAF.
Bias subtraction and sky flat-fields led to satisfactory (flatter
than 1\%) images. The different long exposures were combined with a sigma
clipping algorithm to clean the master exposures of cosmic rays. 

Object search, photometry, and the determination of the completeness
factors were done with the IRAF version of the DAOPHOT II package.
Before applying the final point-source search and PSF fitting, 
we prepared the master exposures in each filter by the following procedure.
First we subtracted from the image the brightest stars and extended objects with
the IRAF {\it imedit} routine. Then we computed an isophotal model (using
the IRAF STSDAS {\it isophote} package) that we 
subtracted from our cleaned image. The shells, dust lanes and ripples were then
visible as residuals. To 
remove also these structures we first subtracted all point-source objects
by the standard DAOPHOT II PSF fitting. Then we applied a 11 $\times$ 11
pixel median filter to the image. The sum of the isophotal model and the median
image was subtracted from the original master exposures, leading to a flat 
background image ready for the final point-source search. We checked that no
systematic offset was introduced in the photometry by this procedure. Due to
model fitting residuals and the very complex structure in the center
of the galaxy we did not consider the innermost 30\arcsec{} radius. 

The calibration was done via typically 40 standard stars from the Landolt
(1992) list by which we transformed our colors to Johnson $V$ and Cousins $I$. 
The calibration agreed well with the aperture photometry values published by
Poulain \& Nieto (1994). 
The mean residuals for $V$ and $I$ are 0.014 and 0.015 respectively. 
Typical errors in color at the faintest magnitudes are about 0.19 mag, about 
0.14 mag at $V\simeq 24.0$, and drop below 0.10 mag for $V<23.5$.

The determination of the completeness was done by standard artificial star 
experiments. The completeness starts to
affect the results at magnitudes of 24.0 mag in $V$ and 23.0 mag in $I$
for the NGC 5018 field and 23.8 mag in $V$ and 22.6 mag in $I$ for the background
field. The limiting magnitudes where the completeness factor has dropped down
to 50 \% (we used this limit for our investigations) are 
24.9 mag in $V$, 23.7 mag in $I$ for the galaxy field, 24.4 mag in $V$ and 23.3 
mag in $I$ for the background field.
%

\section{Colors}

In this section we derive the mean colors of the globular
clusters as well as features in the color distribution.
Background galaxies
can be identified by their second moments of intensities down to $V=23.0$.
In the following we will use the term ``objects'' for objects obeying 
this coarse selection, i.e. evident galaxies 
are rejected.

The color distribution for all the objects in the field detected in $V$ 
{\it and} $I$ is shown in figure
2, where the dashed lines mark our 80\% completeness limits in the $V$ and
$I$ filters.
\begin{figure}
\psfig{figure=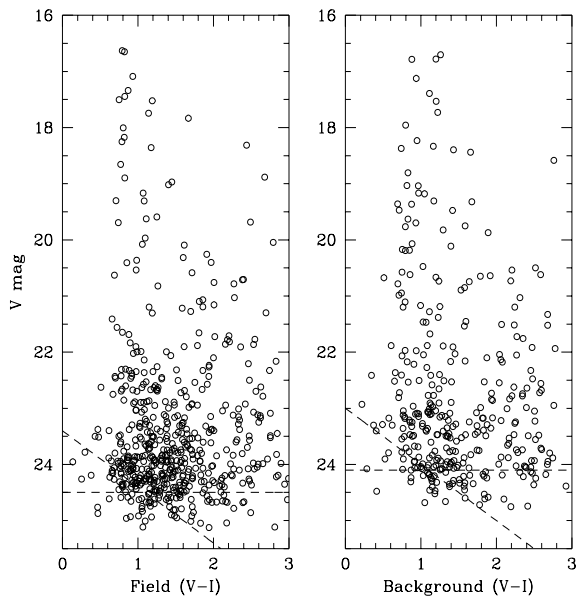,height=8cm,width=8cm
,bbllx=8mm,bblly=57mm,bburx=205mm,bbury=245mm}
\caption{Color magnitude diagram for all objects found in $V$ {\it and} $I$ 
in the
field centered on NGC 5018 (left panel) and in our background field 
(right panel). Note
that the galaxy field is deeper by 0.4 mag in $V$ and $I$ than the background.
The dashed lines mark the 80\% completeness limits in $V$ and $I$.
}
\end{figure}

\subsection{Existence of a dust lane?}

We plotted the spatial distribution of objects of different colors. 
The distribution of blue, $(V-I)<0.8$, objects as well as red, $1.5<(V-I)<2.0$,
objects is shown in figure 3, together with a contour plot of NGC 5018.
The first group is chosen to be bluer
than most background objects, and the second to be redder than most
globular clusters.
\begin{figure}
\psfig{figure=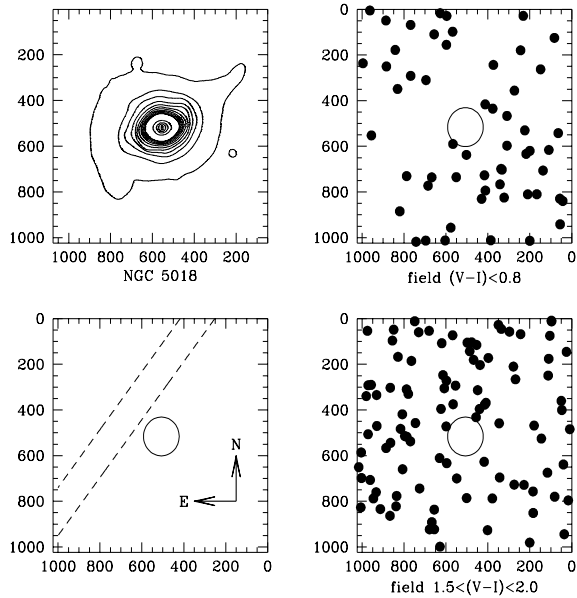,height=8cm,width=8cm
,bbllx=8mm,bblly=57mm,bburx=205mm,bbury=245mm}
\caption{Spatial distribution for objects in the field with $(V-I)<0.8$
(right upper panel), and $1.5<(V-I)<2.0$ (right lower panel), as well as
a contour plot of NGC 5018 (left upper panel). The stripe extending
from the South-East corner to the North of the galaxy, showing no blue
objects and an overabundance of red objects, is indicated in the lower left 
panel. The x and y axis are labeled in pixels of 0.38\arcsec, i.e.~1000
pixels are equivalent to 6.3\arcmin.
}
\end{figure}
Two features appear in these plots: blue objects are totally missing in
a stripe extending from the South-East corner to the North of the galaxy
(0 objects out of 57, in a lane representing 18\% of the surface),
while red objects are overabundant in this stripe (34 objects out of 125,
or 27\% of the objects on 18\% of the surface).

We checked for any flatfield or reduction effect, as well as
any similar feature in the background field, but could not find any.
We exclude the possibility that the inhomogeneous distribution of objects of
different colors could be due to random fluctuations in the spatial 
distribution. The probability for a configuration as we see it is very
small, since the ``blue'' and ``red'' distributions shown above
represent already 3 and $2\sigma$ fluctuations on their own, and are 
correlated to each other as well as with the HI distribution as shown
below. 
We are thus confident that this effect is real, and see two possible
explanations:\\
$\bullet$ different populations of globular clusters have an inhomogeneous 
distribution in NGC 5018. Or,\\
$\bullet$ there is a strong reddening along this stripe, e.g.~a dust
lane.

The first hypothesis is supported by various facts.
As will be shown below, an unrealistic assumption for the gas-to-dust 
ratio would have 
to be made to reconcile the HI observations with the existence of
a dust lane. Moreover, the possible extinction could not be clearly seen in 
the galaxy light on our color image (note however, that this is not trivial,
considering the amount of structure surrounding NGC 5018 and the fact that
it occurs where the surface brightness drops).
Also, the inhomogeneous spatial distribution of the objects 
could be a remnant of the merger event: 
Guhathakurta et al.~(1989) noticed that the HI bridge connecting NGC 5022
and MCG~03$-$34$-$013 seems to bifurcate at approximately the position and
radial velocity of NGC 5018, and while the northern part continues
unbroken, the southern portion disappears near the elliptical.  
This could have lead to an inhomogeneous distribution of
the globular clusters, if for example they formed from the infalling gas in
the southern region, where indeed the blue objects are seen. However
this would not explain the over-density of red objects in the northern
part.

The second hypothesis, the presence of a dust concentration,
is supported by the following points.
The large HI bridge between NGC 5022 and MCG~03$-$34$-$013 
(Guhathakurta et al.~1989), crosses our frame {\it exactly} along the stripe
suspected to be dust (see their Figure 1).
However the column density in the bridge is around several $10^{19}~{\rm atoms~cm}^{-2}$
(Guhathakurta priv.com.) which, using the hydrogen column
density to color excess ratio in our Galaxy of $N_{\rm HI}\simeq5\times
10^{21} E_{B-V} {\rm atoms~cm}^{-2}$ (Bohlin, Savage and Drake 1978), would give an
reddening of $E_{B-V}=0.005$, far too small to produce the effect of
several tenths of magnitudes in $E_{V-I}$ observed, unless the gas-to-dust
ratio is higher by about two orders of magnitude (i.e.
$M_{\rm gas}/M_{\rm dust}=$1 to 5) than in our Milky Way.

Further, emission at $\lambda 60~\mu$m and $\lambda 100~\mu$m at an intensity of
about 1 Jy
was detected with the IRAS satellite. The derived temperature of the infrared
emission was below 30~K, leading to the assumption that this
elliptical galaxy possibly contains cold interstellar gas (Knapp et
al.~1989). 
Unfortunately, even the HIRAS maps (processed IRAS images for a higher
resolution of 1 arcmin) at $\lambda 60~\mu$m and $\lambda 100~\mu$m do still
not have the required resolution to locate the dust in NGC 5018.

Finally, the isophotal shape and the distribution of ionized gas and
dust have been extensively studied by Goudfrooij et al. (1994a, 1994b)
and Goudfrooij \& de Jong (1995).
They noticed that the overall visible optical extinction is about 1
magnitude smaller than expected from the high IRAS flux, and might be
caused by a diffuse distribution of dust, probably accreted from other
galaxies.

We have at this stage no reason to prefer one scenario to the other and will
have to wait for ISO observations in order to solve the ambiguity. 
Nevertheless, we will show that most conclusions are not severely affected by
the assumption of one or other scenario, but that in details our results might
be affected by unknown extinction being clumped or diffuse.

\subsection{Color distributions}

We present in the following the color distributions for both hypotheses. For the
first, we left all colors unchanged. For the second, we made a coarse correction
for reddening within the stripe shown in Fig. 3.
\subsubsection{The sample}
To determine the color of {\it globular clusters}, we selected a sample of 
objects in which the globular clusters dominate the background and compared
this sample with an equally selected one on the background field.
This sample consists of all objects in a elliptical
ring from $30\arcsec$  to $170\arcsec$  major-axis, the lower limit protecting
against variation in completeness towards the center, and the upper
limit being chosen from the density profile: out to a semi-major axis of
170\arcsec, globular clusters dominate the background (see Sect. 5.1).

Further, only objects fainter than $V=22.5$ mag are considered. This is
motivated by the globular cluster luminosity function (see Sect. 4.1),
which shows that globular clusters are only to be expected at fainter
magnitudes. 
\subsubsection{Correction for eventual local extinction}
For the color correction in the stripe we assumed that the colors outside the 
stripe are unaffected. We corrected the colors within the stripe by a constant 
value to match their mean with the mean color outside the stripe.
In the stripe, the objects following the selection described above
have a mean color $(V-I)=1.68$ mag ; in the rest of the ring, the mean color
is $(V-I)=1.38$ mag. We therefore adopted a ``reddening correction''
of $\Delta(V-I)=0.3$ mag, $\Delta V=0.55$ mag , and $\Delta I=0.25$ mag.
In the following, colors and magnitudes in the stripe will be corrected by these
values when we mention a correction for the ``dust lane''.
\subsubsection {The color distribution}
We plotted the color distribution for objects from $V=22.5$ to 24.5 mag
in Figs. 4 and 5. Figure 2 shows that
completeness corrections start to be necessary at magnitudes fainter than
$V=23.5$, and
that the correction is a function of the colors of the objects.

We divided our sample into four categories: $(V-I)=0.5$ to 1.0, 1.0 to 1.5,
1.5 to 2.0, and 2.0 to 2.5. We then applied the correction as a function of
magnitude and of color to the bins in a given category, computed as the
product of the mean $V$ completeness and the corresponding mean $I$ 
completeness\footnote{
We examined in artificial star experiments the effects of combining
completeness factors for $V$ and $I$. In our case the completeness drops
due to photon noise rather than to crowding, so that completeness in $V$
and $I$ are independent.
We therefore used the product of the completeness factors in $V$
and in $I$, rather than just the lowest completeness at a given
magnitude.}.
See Tab. 2 for the values used for the field and
the background.
\begin{table}
\caption{Completeness values in percent as a function of magnitude and color for
the field (left value) and the background (right value).}
\begin{tabular}{c c c c c}
\hline
Mag. / $(V-I)$ & 0.5---1.0 & 1.0---1.5 & 1.5---2.0 & 2.0---2.5 \\
\hline
23.0--23.5 & 95 95 & 100 100 & 100 100 & 100 100 \\
23.5--24.0 & 87 77 & 95 95 & 100 100 & 100 100 \\
24.0--24.5 & 68 22 & 87 54 & 88 66 & 91 68 \\
\hline
\end{tabular}
\end{table}
The color distribution of the background was smoothed by the median over 
respectively 3 bins to avoid any large fluctuations due to galaxy clusters
that we could identify on our background frames, which fall
in single color bins. It was then scaled down to the ring
area considered in the galaxy field and subtracted bin by bin from the color 
distribution of objects in the ring around NGC 5018.

\begin{figure}
\psfig{figure=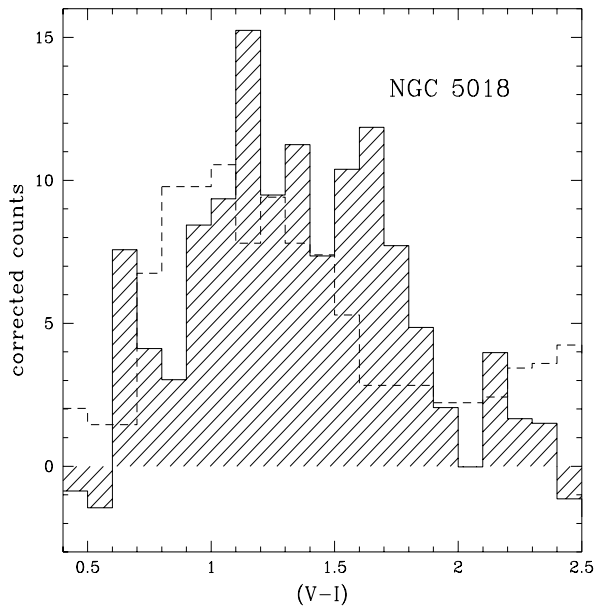,height=8cm,width=8cm
,bbllx=8mm,bblly=57mm,bburx=205mm,bbury=245mm}
\caption{Color distribution of globular cluster candidates between 30\arcsec and
170\arcsec, with magnitudes $22.5<V<24.5$. 
The histogram shows the final corrected counts, the
histogram of the background used for correction is over-plotted (dashed
line).  No correction for the ``dust'' lane was applied.
}
\end{figure}
\begin{figure}
\psfig{figure=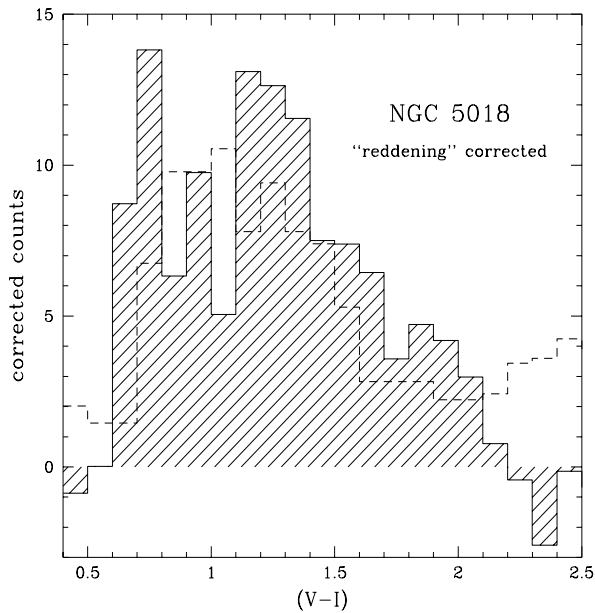,height=8cm,width=8cm
,bbllx=8mm,bblly=57mm,bburx=205mm,bbury=245mm}
\caption{Color distribution of excess objects between 30\arcsec and
170\arcsec, with magnitudes $22.5<V<24.5$. 
The histogram shows the final corrected counts, the
histogram of the background used for correction is over-plotted (dashed
line). The colors in the stripe (see figure 3) were corrected by
$\Delta(V-I)=0.3$ mag. 
}
\end{figure}
The resulting color distribution is much broader than expected from the 
photometric errors alone. If all clusters were concentrated at a single color,
we would obtain a gaussian distribution with a FWHM of about 0.4 mag. 
The distribution uncorrected for reddening shows different groups of objects:
a small group of objects around $(V-I)=0.7$ mag,
a broad population around $(V-I)=1.2$ mag, and a narrower one around 
$(V-I)=1.65$.

The color distribution corrected for the reddening in the ``dust lane'' shows 
a different mixing of the populations. The population at $(V-I)=0.7$ becomes as
large as the one at $(V-I)=1.2$, and the population at $(V-I)=1.65$, shifting
to a somewhat bluer color ($(V-I)=1.50$), is 
only seen as a tail of the population at $(V-I)=1.2$.

Both distributions show some red objects clustered around
$(V-I)=2.2$ or $(V-I)=1.9$ respectively. In the case of no reddening, they
would be a galaxy cluster at low redshift, judging from their color. In
case of a reddening stripe the whole group shifts towards the blue:
all these objects appear to lie in the stripe.
This shows that our reddening correction of 0.3 mag might be very crude, and
that, in fact, some objects could be even more reddened.
\subsection{Interpretation of the color distribution}
\subsubsection{The colors}
In the following we include
the Galactic extinction of $E(B-V)=0.053$ mag or $E(V-I)=0.072$ mag 
towards NGC 5018 (Burstein \& Heiles 1984).

If we compare the colors of the globular clusters in NGC 5018 with those
of the Galactic globular clusters (see Fig. 6, data taken from Harris
1996), we notice that the
distribution of the two red populations is similar to that of the
globular clusters in the Milky Way,
if the colors of the latter are not corrected for extinction.
\begin{figure}
\psfig{figure=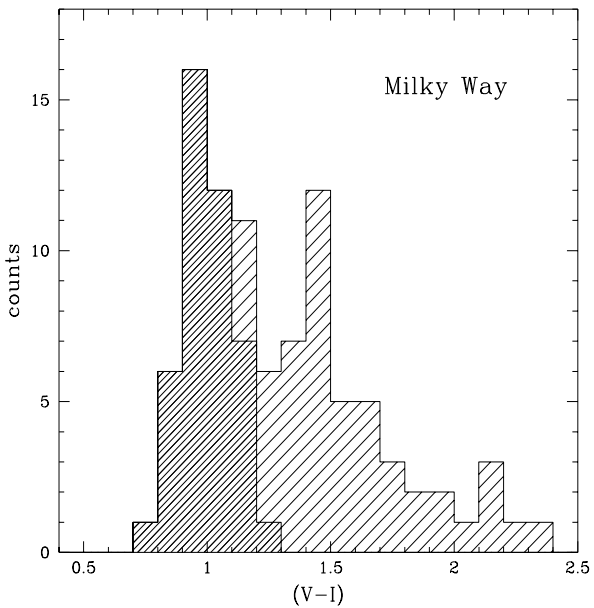,height=8cm,width=8cm
,bbllx=8mm,bblly=57mm,bburx=205mm,bbury=245mm}
\caption{Color distribution of the Galactic globular clusters,
uncorrected for reddening. The finer hashed part is the contribution of
the globular clusters for which the reddening $E(B-V)$ is less than
0.2. Data from Harris (1996).}
\end{figure}
From their colors, the red population(s) are thus compatible with being
old globular clusters as seen in the Milky Way, with metallicities
ranging from $[Fe/H]=-2.5$ to $0$ dex and possibly suffering from as
much reddening. We cannot discriminate between
single populations with a wide range of metallicity combined with some
reddening due to the dust in NGC 5018, or eventually two distinct
populations of old globular clusters.

However, the blue objects are bluer than even the most metal-poor
globular clusters in the Milky Way. Therefore they must be younger, unless 
they are more
metal-poor than any observed globular cluster to date (e.g.~Geisler et
al.~1995).
According to the models from Fritze v.Alvensleben \& Burkert (1995),
agreeing with Bruzual \& Charlot's (1993) model for solar metallicity,
the globular cluster population centered at $(V-I)_0=0.63\pm0.10$ (eventually
redder by 0.1 in the corrected case), could have an age ranging from 200 Myr 
to 6 Gyr, depending on the metallicity of the globular clusters (the
younger limit being for extremely metal rich ($Z=4\times10^{-2}$), the
upper for extremely metal poor ($Z=1\times10^{-4}$) globular clusters).
This compares well with the young globular clusters found in NGC
7252 (Whitmore et al.~1993), which are marginally bluer (by about 0.1
mag) and brighter by 2--3 magnitudes than our candidates. And thus probably
younger by a few 100 Myr. These clusters are estimated to have ages between 
0 and 800 Myr 
(Schweizer \& Seitzer 1993).

All this supports the fact, that the young  globular clusters in NGC
5018 could have formed 
during the last interaction dated between several hundred Myr and
a few Gyr ago (see references in Sect. 1).
\subsubsection{The color distribution with magnitude}
In Table 3 we show the counts of globular clusters (background
subtracted) for two magnitude bins, in order to examine the distribution
of the red and blue objetcs with the magnitude.
\begin{table}
\caption{Number of globular clusters after background subtraction in different
color intervals (see
Figs. 4 to 5 for the histogramms) for the case where no dust lane is
taken into account, and for the case where coorections are made for a
dust lane as explained in Sect. 3.2.2.}
\begin{center}
\begin{tabular}{l r r r}
\hline
color interval: & 0.6---0.9 & 0.9---1.4 & 1.4---1.9 \\
mag limit & & & \\
\hline
\multicolumn{4}{c}{uncorrected case}\\
22.5---23.5 & 1.4 & 13.6 & 13.2 \\
22.5---24.5 & 13.2 & 56.8 & 42.3 \\
\hline
\multicolumn{4}{c}{corrected case}\\
22.5---23.5 & 4.5 & 12.7 & 7.6 \\
22.5---24.5 & 28.8 & 52.1 & 29.7 \\
\hline
\end{tabular}
\end{center}
\end{table}
For the two red intervals the counts increase by a factor 4 when 
going a magnitude deeper 
(independently of a correction or not).
Assuming a turnover magnitude around $m^0_V = 26.3$ mag (see Sect.4.1), 
we cut off respectively 2.8 and 1.8
mag below the turnover magnitude and sample respectively 3\% and 12\% of
the surface of the globular cluster luminosity function (assuming it to
be Gaussian and to have a dispersion of 1.4 mag). This is in good 
agreement with the increase by roughly a factor 4 found for the 
red populations.

For the blue objects, the statistics are too uncertain to be used in a
quantitative way. Qualitatively, we detect only a few (1--5) objects down to
$V=23.5$ (including also the interval $V=20.5$ to 22.5, i.e.~$M_V
\geq -13$, adds 3 more objects) 
and then a sudden increase (13--30) when sampling a magnitude
deeper. Assuming an unimodal luminosity function for the blue objects,
not much broader than the luminosity function of red objects, this would
hint at a peak about 2--3 magnitudes brighter in $V$ than that of the red
objects.
This fits into the scenario where the blue objects would indeed be
younger:  
according to e.g.~Fritze v.Alvensleben \& Burkert's model, globular cluster
around 1 Gyr would be 3 magnitudes brighter in $V$ than 15 Gyr old ones.

To get an estimate of the ratio of blue to red globular cluster candidates we 
assume a ``blue GCLF'' with a Gaussian shape, peaking 2 to 3 magnitudes
brighter than the GCLF of the red objects, for which we take the
turn-over values from table 4. This leads then to 20--150 blue globular
clusters in total, or 1--10\% of the red objects (see Sect. 4.2).

In summary, despite the fact that we have small number statistics
for the globular clusters and large uncertainties in the dust
distribution, the following points can be retained:\\
$\bullet$ A blue population of globular clusters exists in NGC 5018, and is 
probably between 200 Myr to 6 Gyr old. 
(the upper range would be slightly preferred from their magnitudes, when
compared to the young clusters in NGC 7252).
These young globular clusters are likely to have 
formed in the last interactions within the galaxy group.\\
$\bullet$ The red globular cluster population(s) is compatible with old 
($\geq$ 10 Gyr) globular clusters.
Two distinct populations might exist, but the data are 
still compatible with a single population differentially reddened and
with a large spread in metallicity.
%
%
\section{The number of globular clusters}

The globular cluster luminosity function (GCLF)
plays an important role as extragalactic distance indicator and is
essential for the estimation of the total number of globular clusters and
specific frequencies.
In recent years a lot of work has been done towards understanding
the representation of the GCLFs and the fine tuning with the dependences on
metallicity and age (e.g. Secker 1992, Ashman et al. 1995).

\subsection{The globular cluster luminosity function}

To compute the GCLF, we used the
coarse selection of the objects described in the Sect.
3.~(i.e.~obvious galaxies were rejected)
and restricted ourselves to an elliptical ring from 53\arcsec{} to 170\arcsec{}
major-axis. We made no selection in color, since we wanted to avoid a combined
completeness correction. Thus we included the blue objects and one has to
be aware that these objects may influence the following results as discussed
later. 

To determine the GCLF,
we counted the selected objects in bins of 0.5 mag width and
shifted the bins in steps
of 0.125 mag to get four luminosity distributions. We then
corrected the counts for incompleteness. These steps were done for the field
centered on NGC 5018 as well as for the background field.
We interpolated the background luminosity function
by a polynomial fit of the 4th order
to smooth out the effect of individual background galaxy clusters
biasing given bins. This background luminosity function was then
subtracted from the field luminosity function bin by bin.

The limiting magnitude of 24.4 mag in $V$ is determined by the last bin
of the background luminosity function with a completeness  above 50\%.
Since we only have 7 to 8 data points for
each fit to the bright tail of
the luminosity function, 
we combined respectively the two luminosity functions shifted by 0.25
mag to obtain distributions with 0.5 mag bins in 0.25 mag
steps. The points are then no longer statistically independent, but
stabilize the fit.

The best representation of GCLFs are Gaussians or Student $t_5$ functions
with dispersions of around $\sigma_{\rm Gauss} = 1.4$ or $\sigma_{t_5} = 1.1$
respectively (e.g.~Secker 1992).
The resulting turnover values (error-weighted means) from fitting 
these functions to the double-binned distributions
vary a lot as a function of the chosen dispersion, while being almost unaffected
by the correction for the dust lane or different binnings. In case of a 
Gaussian fit the values range
from $V_{TO}=25.9$ for the lowest dispersion of $\sigma_{\rm Gauss}=
1.2$ to $V_{TO}=26.9$ mag for the highest dispersion of 
$\sigma_{\rm Gauss}=1.5$. For a fit with a $t_5$ function the values range from
from $V_{TO}=25.7$ for the lowest dispersion of $\sigma_{t_5}=
1.0$ to $V_{TO}=26.4$ mag for the highest dispersion of
$\sigma_{t_5}=1.3$.  
The turnover values have been corrected for a foreground extinction
of $A_V = 0.17$ mag (Burstein \& Heiles 1984). 

The uncertainties in our results are the following:\\
-- The accuracy of the turnover determination by a given
function lies between 0.3 and 0.6 mag for a sparse sample like ours, 
covering only the bright end of
the GCLF. This was derived from Monte Carlo simulations, simulating
samples similar to ours and fitting them in the same way. Single deviating 
bins can largely influence the fit.\\
-- Figure 7 shows that the fits of a Gaussian and a $t_5$ function 
with different dispersions are 
indistinguishable from each other in the fitted bright tail, whereas the
turnovers cover a range of about one magnitude. Since we do not know the real 
dispersion for the GCLF of NGC 5018, we can only derive turnover values within 
this range. Nevertheless, in the case of a broad color distribution one would 
expect a dispersion around $\sigma_{\rm Gauss} = 1.4$, as seen for
example in M87 which has a comparable color distribution (Whitmore \& 
Schweizer, 1995).\\
-- A neglected young globular cluster population could affect the
bright tail of the luminosity function. 
The possible young population which is 2--3 mag
brighter than the old one would produce a more 
extended, flatter wing on the bright side of the GCLF 
which leads to 
underestimation of the turnover values for the old population. 
Monte Carlo simulations showed that in our case (less than 10\% young
clusters and only fitting the bright tail) the underestimation would be less
than 0.1 mag, but would introduce an additional scatter of 0.5 mag.
\begin{figure}
\psfig{figure=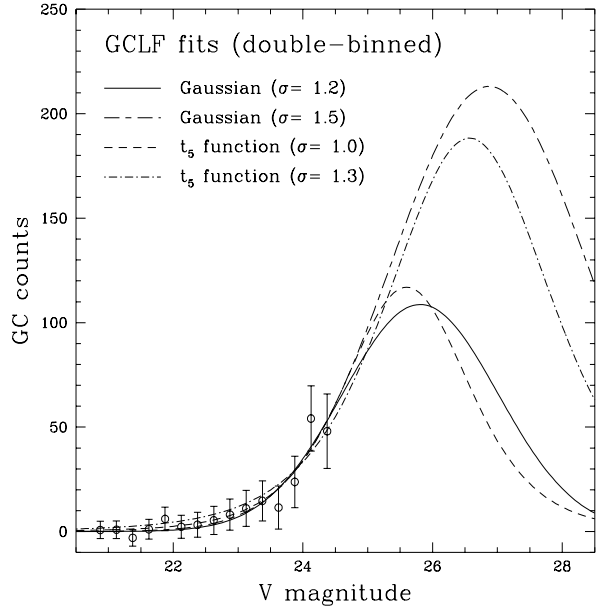,height=8cm,width=8cm
,bbllx=8mm,bblly=57mm,bburx=205mm,bbury=245mm}
\caption{The globular cluster luminosity function of NGC 5018 
 fitted by Gaussians and $t_5$ functions with different dispersions 
(see Sect. 4.1). In the bright tail all of the fits are equally good}
\end{figure}

Including these uncertainties we can estimate a distance for NGC 5018 using 
the GCLF. We assume an absolute
turnover value corrected for metallicity of $M_V^0 = -7.1 \pm 0.2$ mag
(following Ashman et al.~1995, combined with a metallicity slightly less
than solar for our old population, assumed from our colors).

Our average values 
lead to a distance modulus in the range of $(m - M)_V^0 = 32.8 \pm
0.4$ to $(m - M)_V^0 = 34.0 \pm 0.6$, bracketing the value given in 
Tully (1988), derived from the heliocentric velocity of $v = 2794 
\pm 15~{\rm km~s}^{-1}$, 
and assuming $H_0=75~{\rm km~s}^{-1}~{\rm Mpc}^{-1}$. 

An accurate distance determination by the GCLF awaits deeper data.

\subsection{The total number}

The calculation of the total number of globular clusters strongly depends
on the fraction of the GCLF that we have covered with our data.
The fitted luminosity function covers all clusters in the ring of 53\arcsec{}
to 170\arcsec{} of major-axis. The results of the Gaussian fits of Tab. 4 (with
``dust'' lane stripe) range between $500 \pm 200$ and $1000 \pm 600$ objects.
For the inner and outer parts one has to make assumptions for the number
of globular clusters
based on the radial distribution of the globular cluster density (see section
5.1). Due to disruption processes we do not expect a steep increase of
the cluster density in the
innermost 10 kpc from the center but
a flattening of the density profile 
(e.g.~McLaughlin et al. 1993).
Therefore, a reasonable guess
for the inner ring (0\arcsec{} - 53\arcsec{}) is a density of $29.3 \pm 5.0$
objects per square arcmin (the density at 40\arcsec{}) leading to $50 \pm 9$
objects down to a magnitude of $V = 25$ mag.
For the outer part (170\arcsec{} - 200\arcsec{}) we assumed a mean density of 
$4.0 \pm 1.7$ leading to $27 \pm 12$ objects.
We then corrected for the full GCLF. 
Our estimate of the total number of globular clusters in NGC 5018 ranges
from $800 \pm 270$ to $1700 \pm 750$.

\subsection{The specific frequency}

The specific frequency is defined as $S = N\times 10^{0.4(M_V+15)}$,
where $N$ is the total number of globular clusters in the galaxy and $M_V$ its
absolute visual brightness (Harris \& van den Bergh 1981). Mean values
for $S$ are found to correlate with the galaxy type, luminosity, and  
environment.
While the mean specific frequency of spirals is around 1, ellipticals in
small groups have a higher mean value around 2.6, and ellipticals in clusters
have a mean value around 5.4 (Harris 1991).
Kumai, Hashi, and Fujimoto (1993) tried to quantify empirically the
correlation of the specific frequency
$S$ with the local galaxy density. For the given galaxy density around NGC 5018
of 0.29 galaxies/Mpc$^3$ (Tully 1988), their relation would lead to a value of
$S = 3.0 \pm 0.2$, also expected from Harris' (1991) compilation. Djorgovski
\& Santiago (1992), supported by Zepf, Geisler \& Ashman (1993) showed the 
relation between galaxy luminosity and number of globular cluster to be 
$N_{gc}\propto L^2$ rather than $N_{gc}\propto L^1$, from which we also would 
expect an $S$ value around 3 for NGC 5018.  

The absolute visual brightness of NGC 5018 depends on its distance, but
since the derived total number of clusters, taken from the GCLF, also increases
with distance, the calculated $S$ values
lie in a narrow range between $S = 0.9 \pm 0.5$ and $S = 1.1 \pm 0.6$,
significantly below the expected value of $S = 3.0$ for our given $N$ and
$M_V$ values. For NGC 5018 to fall in the mean for
ellipticals in small groups this would mean: either having 3 times as many
clusters as observed, or being roughly one magnitude ``too bright''.
One explanation could be that the absolute luminosity of NGC 5018 is
increased by a young stellar population and therefore $S$
is underestimated. 

We calculated the increase of the luminosity
and the change in color when mixing a young blue and red old population
based on the models given by Fritze v.Alvensleben \& Burkert (1995).
If one wants to reach a $S$ value of 3 
the young population would have to have increased the total 
$V$ luminosity of NGC 5018 by about 1.1 mag. 
A 3 Gyr old population with a color of $V -I = 0.75$ mag which contributes
33\% to the total $V$ luminosity would
increase the luminosity by that amount. The color would change from 
$V -I = 1.25$, which is 
usual for ellipticals (e.g. Goudfrooij et al. 1994b), to $V -I = 1.05$, 
the color of NGC 5018.

\section{Spatial distributions}

\subsection{The radial distributions}

In the following we derive the radial density profile of the globular clusters
and compare it with the profile of the galaxy light.

We compute the radial profile using all the objects found on our frame
down to a magnitude of $V=25.0$. The objects were binned into elliptical
rings 100 pixels (37.7\arcsec) wide, the counts were then corrected for
finding incompleteness (using 100\% down to $V=24.0$, 94\% between
$V=24.0$ and $24.5$, and 65\% between $V=24.5$ and $25.0$), and
geometrical completeness when the ring expanded over the limits of our frame
(only necessary for the ring 500 to 600 pixels ($189\arcsec$ to $226\arcsec$)).

Table 4 shows the mean semi-major axis of the ring, the corrected
counts, the applied
geometrical completeness, the total ring area, and the density. The
density error was taken as a combination of the square root of the raw counts
and uncertainties in the correction for incompleteness. The second
part of Table 4 shows the object densities on the background field, where
no rings were defined, but 3 stripes perpendicular to the minor axis.
\begin{table}
\caption{Radial counts of objects around NGC 5018 (upper part) and in
the background field (lower part)}
\begin{tabular} {r l l l l}
\hline
radius & corrected & geom.  & full area & density \\
in arcsec & counts & corr. & in arcmin$^2$ & obj/arcmin$^2$ \\
\hline
56.5 & 105.6  & 1.00 &  2.61 &  $40.5\pm 3.9$ \\
94.3 & 121.4  &  1.00 &  4.34  &  $28.0 \pm 2.5$  \\
132.0 &  148.0 &   1.00 &   6.08 &  $24.3 \pm 2.0$ \\
169.7 &   200.2 &  0.96 &  7.82 &  $25.6  \pm 1.8$ \\
207.4 &   182.4 &   0.79 &  9.56 &  $19.1 \pm  1.6$ \\
\hline
290.3  & 280 & 1.0 & 14.7 & $19.0\pm2.5$ \\
444.9  & 250 & 1.0 & 13.3 & $18.8\pm2.5$ \\
662.8  & 254 & 1.0 & 13.3 & $19.1\pm2.5$ \\
\hline
\end{tabular}
\end{table}

On the background field, the counts at $V=24.5$ mag are already
incomplete by more then 50\%. We therefore corrected these counts by
comparing them with Smail's et al.~(1995) deep counts in $V$, and extrapolated
our results down to $V=25.0$ mag with their slope. We find for our 
background field a mean density of $19.0\pm1.2$ objects per square
arcmin down to $V=25.0$ mag, in agreement with Smail's et al.~(1995)
results.

The radial profile is plotted in Fig. 8. 
We fitted a power-law of the kind $\rho \sim r^{-x}$, where $\rho$ stands
for the surface density of the globular clusters or the light flux of the 
galaxy, and
$r$ for the semi--major axis, to both globular cluster and light profiles in
order to compare their fall-off.

For the density profile of the globular clusters after
subtraction of the background density, we obtain $x = 1.3 \pm 0.4$ (the
large error being due to the weight attributed to the fourth point). The
fit to the light profile (between $10\arcsec$ and $160\arcsec$ semi-major axis)
returns a slope of $x= 2.4\pm 0.1$, and is therefore steeper than
the globular cluster profile. This last slope depends slightly on how
far one extends the fits, since the galaxy light gets steeper beyond a
semi-major axis of about $100\arcsec$. Stopping our fit there returns
$x= 2.2\pm 0.1$, essentially the same value that would return a
fit to the Goudfrooij et al.~(1994a) data in this range, for which we find
$x= 2.15\pm 0.05$.
\begin{figure}
\psfig{figure=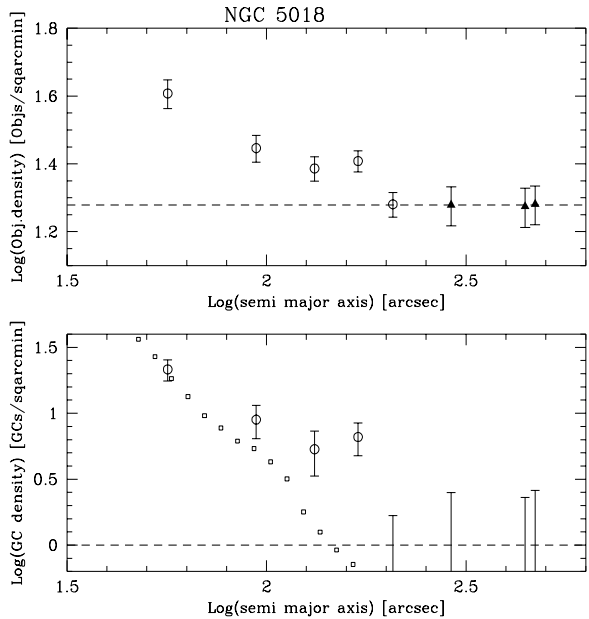,height=8cm,width=8cm
,bbllx=8mm,bblly=57mm,bburx=205mm,bbury=245mm}
\caption{Radial density profile of the globular clusters. The upper
panel shows the raw object densities in counts/arcmin$^2$ down to
$V=25.0$ mag (circles are from the
field, triangles from the background); the dashed line marks our mean
object density on the background field. 
The lower panel shows the background corrected densities, i.e.
the density of the globular clusters; here we over-plotted with an
arbitrary offset the galaxy light profile (squares).
}
\end {figure}

The galaxy light does not have an unexpectedly steep profile, it is rather
the globular cluster system which extends further out
and is flatter than the stellar component. A similar behavior was
already observed in galaxies with rich globular cluster systems, while
in sparser systems the globular clusters seem to follow the galaxy light
(e.g.~Kissler-Patig et al.~1996 and references therein).

\subsection{Ellipticity and position angle of the globular cluster
system}

\begin{figure}
\psfig{figure=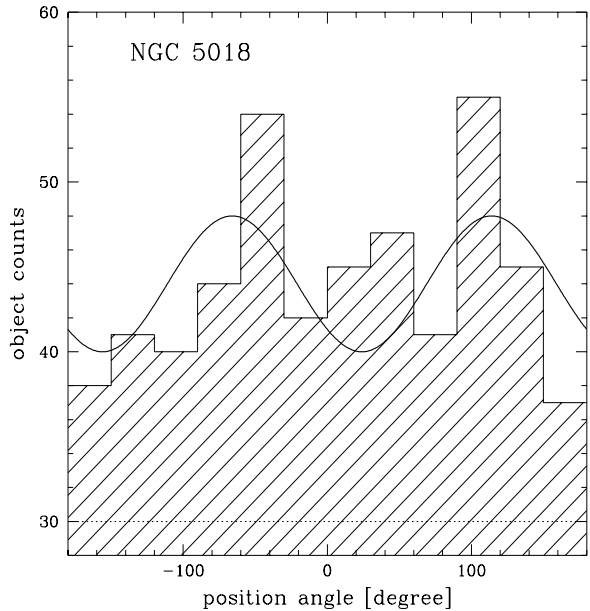,height=8cm,width=8cm
,bbllx=8mm,bblly=57mm,bburx=205mm,bbury=245mm}
\caption{Angular distribution of the objects around NGC 5018. The dashed
line shows the expected background level, the solid curve is the best
fit of a double cosine to the data.}
\end{figure}
Figure 9 shows the angular distribution of all objects down to a magnitude
of $V=25$ around NGC 5018 in a ring from $38\arcsec$ to $188\arcsec$,
uncorrected for finding incompleteness; that we assumed to be homogeneous over the
position angle. 
Objects closer than $38\arcsec$  were not considered because of the too high
finding incompleteness within this radius.
To avoid any geometrical
incompleteness, we limited the counts to $188\arcsec$. 
The thus defined ring area 
covers 29.6 arcmin$^2$, or 73\% of the total frame surface.

We divided the ring into 12 segments, 30 degrees wide each, and counted the 
objects in the segments. 
Each ring has a surface of 2.48 arcmin$^2$.
We estimated the background contamination, uncorrected for completeness,
from the area outside $190\arcsec$, where the density profile reaches
already the background (see Sect. 5.1.) and obtained 12.0 objects per square
arcmin.  
We thus expect 30
objects in each segment (marked as a dashed line in Fig. 9). 

The two strong peaks in Fig. 9
separated by 180 degrees at the position angle of the galaxy
hint at a non-spherical distribution of the objects. Indeed, the distribution
can be better fit by a double cosine (expected for an elliptical
distribution) than by a constant function (expected for a purely
spherical distribution); note however that a spherical distribution is
still compatible with the data within $2\sigma$.

The best fit by a double cosine of the form $n = A*cos(2*(\phi - B)) + C$, 
where $n$ are the corrected counts and $\phi$ the position angle E of N, 
returns the parameters $A=4\pm2$, $B=114\pm25$, and $C=11\pm2$.
The globular cluster system thus has a position angle of
$114\pm25$ degrees E of N, and its ellipticity as $1-(C-A)/(C+A) = 0.5
\pm 0.3$.  The isophotes of the galaxy have a position angle of $94 \pm
2$ out to $30\arcsec$ and then become more irregular, shifting between
95 and 110 degrees E of N (Goudfrooij et al.~1994b). The ellipticity of
the isophotes is roughly $0.3 \pm 0.1$ (Goudfrooij et al.~1994b), while
the best oblate model for the galaxy light returns 0.6 (Carollo \&
Danziger 1994).

The globular cluster distribution is compatible within the errors with
the galaxy light. It seems slightly shifted with respect to the stellar 
light and would rather support Carollo \& Danziger's (1994) model where the
galaxy is intrinsically flat (E4 to E6) but seen at an inclination of
30 to 55 degrees, if we assume that the globular clusters share the
ellipticity of the luminous component.
%

\section{Discussion and Conclusions}

\subsection{Verifying the prediction of globular clusters in {mergers}}

At least some ellipticals are supposed to have formed by mergers or
interactions. When the majority of these events happened is still
unclear, as well as the evolutionary state of the progenitors (still
gaseous proto-galaxies or evolved stellar systems?, see e.g.~ White (1994) for 
a review). 

NGC 5018 is probably one of the elliptical galaxies that were formed
by mergers or interactions of evolved stellar systems. Guhathakurta et
al.~(1989) suggest an encounter of at least three bodies including at least
one cold disk system.

The properties of the globular cluster
system of NGC 5018 can be compared with the predictions of the model for 
globular cluster formation in mergers (Ashman \& Zepf 1992 and Zepf \& Ashman 
1993). Several predictions of the models are confirmed:

(i) -- NGC 5018 most probably contains ``young'' (several hundred Myr
to 6 Gyr) globular clusters that could be attributed to the last
interaction of NGC 5018. This would close the gap between different
``old'' populations in the same galaxy and ``newborn'' globular
clusters, as seen directly in merging or interacting galaxies (e.g.~ NGC
3597: Lutz 1991; NGC 1275: Holtzman et al.~1992; NGC 7252: Schweizer \& Seitzer
1993; He2-10: Conti \& Vacca 1994).

(ii) -- In NGC 5018, the problem caused by age/metallicity ambiguity in
broad-band filters is partly solved by the big differences in
age between the clusters, that make the young population clearly stand
out in color. It remains however very difficult to decide whether one
or two ``old'' populations are present, owing to the large uncertainties
in the dust distribution within the galaxy. Therefore it is not
possible to say whether two similar systems (e.g.~spiral galaxies)
merged, in which case different old populations are expected, or that a larger 
system (already elliptical?) interacted with smaller systems.

Different populations of globular clusters around elliptical galaxies,
as identified in color distributions (e.g. examples in Zepf et al.
1995, or the recent study by Whitmore \& Schweizer 1995 in M87), wait
for confirmation by  spectroscopy of whole globular cluster systems with 
10m-class telescopes.

(iii) -- The model also predicts flatter radial density profiles for globular cluster
after a merger event. The assumption is that old and new globular clusters 
follow the distribution of the field stars with which they
formed. During a merging event the gas condenses to the center to
form the new stars and clusters, so that the resulting stellar
distribution is more centrally concentrated than the old globular
clusters. If, after the interaction, the old globular clusters still
dominate the system, it will appear flatter then the galaxy light. If,
on the contrary, the new globular clusters dominate, the system will only be
marginally flatter then the light of the galaxy.

In the case of NGC 5018, the globular cluster system extends 
further than the stellar component. Therefore, if the assumption above
is right, we would expect only very few globular clusters to have
formed newly and the old globular clusters to still dominate the system, as
is indeed the case (see Sect. 4.1), since we estimate the young
population to be at most 10\% of the old one.

In that respect, however, poor systems that follow the profile of the
galaxy light like NGC 720 (Kissler-Patig et al.~1996) would not fit into
a merger scenario. Density profiles of globular clusters could then
serve to identify ellipticals that did not form by late stellar
mergers.

One prediction of the model, however, does not seem to be verified here:
originally, the model aimed to explain the higher specific frequency
of ellipticals compared with spiral galaxies. Taking into account the different 
mass to $V$ luminosity along the Hubble sequence, the specific frequencies of
ellipticals are higher by a factor two than in spirals (Ashman \& Zepf
1992).
If this excess of globular clusters had to be explained by formation during 
mergers and interactions alone, Ashman \& Zepf (1992) expected the formation 
of as many new globular clusters as there were old ones.

The specific frequency of NGC 5018 is not much higher than in spirals,
mainly due to its high luminosity.
To reconcile its $S$-value with that of normal
ellipticals in groups (about 3, Harris 1991)
the young stellar population must have raised the absolute luminosity of
NGC 5018 by one magnitude.

The results for NGC 5018 show, however, that only few blue globular
clusters formed in this last interaction. We estimate the young population 
of globular clusters
to make up less than 10\% of the total population. At least in this
case, the merger event did not produce a sufficiently high number of new
clusters to systematically explain the higher number of globular
clusters found in ellipticals as compared to spirals. Either NGC 5018
experienced more mergers in the past that were more effective, or
another scenario is needed to explain the systematic higher number of
globular clusters around ellipticals.

Harris (1995) estimated that for an 
efficiency of 1\% to turn gas into bound stellar clusters during a merger,
typical
merging events involving $10^{10}~{\rm M}_\odot$ of gas would produce about 300
clusters with an ``average'' mass of $3\times 10^{5}~{\rm M}_\odot$. 
In the case of NGC 5018 roughly 100 new clusters formed, which might allow
an estimation of the amount of gas involved.
Note that the efficiency seem to be influenced by the amount of gas in 
both merging galaxies. NGC 5018, probably an elliptical merging with a cold
disk system (Guhathakurta et al.~1989), resembles in that respect NGC 5128, 
the product of a bulge dominated system with a late--type spiral (Minniti et 
al.~1996), for which Zepf \& Ashman (1993) suggested that the small amount 
of newly formed clusters was due to the swallowing of a small satellite.
In contrast, systems like NGC 4038/4039 (the ``Antennae''), the result of 
a Sb and a Sc galaxy merging, produced over 700 young globular 
clusters (Whitmore \& Schweizer 1995). This suggests that ``gas on gas'' infall 
creates very favorable conditions for the formation of globular clusters, while
the infall of gas into an early--type galaxy is not as effective.

\subsection{Globular clusters compared to dark matter}
In their kinematical study, Carollo \& Danziger (1994) found through dynamical
modeling that NGC 5018 is much better fit by a model with a dark
matter halo than without. Their best oblate model with dark matter returns 
the presence of a dark halo six times as massive, twice as diffuse, and
half as flattened (about E3) as the luminous galaxy.
In that respect it is interesting to note, that the profile for the globular 
cluster system was computed further than 38\arcsec~($> 2R_e$), where the
dark matter potential will dominate, and is also twice as diffuse as
the stellar component. However, its ellipticity seems higher than that
returned by the dynamical model for the dark halo, even if both still
agree within the errors.

The association of globular cluster systems with the stellar or the
dark halo also waits verification from kinematical studies of entire globular cluster
systems.
%

\acknowledgements
We wish to thank Tom Richtler for his constant support, as well as
Paul Goudfrooij and Edwin Huizinga (``the
unbelievers'') for helpful discussions. Uta
Fritze-v.Alvensleben is thanked for providing us with the tabulated results 
of her globular cluster color evolution computations.
Thanks also to Do Kester for providing the HIRAS images of NGC
5018.
The referee S.~Zepf is acknowledged for useful comments.  
This research was partially supported through the Deutsche Forschungsgemeinschaft
under grant Ri 418/5-1.
This research made use of the NASA/IPAC extragalactic database (NED) which is 
operated by the Jet Propulsion Laboratory, Caltech, under contract with the
National Aeronautics and Space Administration.

\enddocument